\documentclass[twocolumn,showpacs,aps,prl,superscriptaddress]{revtex4}

\usepackage{graphicx}
\usepackage{dcolumn}
\usepackage{amsmath}
\usepackage{epsfig}

\input pubboard/babarsym.tex

\def\leptontag{{\tt Lepton}}
\def\kaonitag{{\tt Kaon\,I}}
\def\kaoniitag{{\tt Kaon\,II}}
\def\othertag{{\tt Inclusive}}

\newcommand{\BABARPubYear}    {02}
\newcommand{\BABARPubNumber}  {008}

\newcommand{\SLACPubNumber} {9293}

\def\figurebox#1#2#3{%
    \def\arg{#3}%
    \ifx\arg\empty
    {\hfill\vbox{\hsize#2\hrule\hbox to #2{\vrule\hfill\vbox to #1{\hsize#2\vfill}\vrule}\hrule}\hfill}%
    \else
    {\hfill\epsfbox{#3}\hfill}%
    \fi}

\begin{document}

\preprint{\babar-PUB-\BABARPubYear/\BABARPubNumber}
\preprint{SLAC-PUB-\SLACPubNumber}

\begin{flushleft}
\babar-PUB-\BABARPubYear/\BABARPubNumber\\
SLAC-PUB-\SLACPubNumber\\
\end{flushleft}

\title{
{\large \bf  Measurement of the {\boldmath\CP} Asymmetry
 Amplitude {\boldmath\stwob}}
}

%
\author{B.~Aubert}
\author{D.~Boutigny}
\author{J.-M.~Gaillard}
\author{A.~Hicheur}
\author{Y.~Karyotakis}
\author{J.~P.~Lees}
\author{P.~Robbe}
\author{V.~Tisserand}
\author{A.~Zghiche}
\affiliation{Laboratoire de Physique des Particules, F-74941 Annecy-le-Vieux, France }
\author{A.~Palano}
\author{A.~Pompili}
\affiliation{Universit\`a di Bari, Dipartimento di Fisica and INFN, I-70126 Bari, Italy }
% ...
\collaboration{The \babar\ Collaboration}
\noaffiliation

\date{\today}

\begin{abstract}
We present results on time-dependent \CP
asymmetries in neutral $B$ decays to several \CP eigenstates. The
measurements use a data sample of about 88 million $\FourS\to B\Bbar$
decays collected between 1999 and 2002 with the \babar\ detector at the
\pep2\ asymmetric-energy \BF\ at SLAC. We study events
in which one neutral $B$ meson is fully reconstructed in a final state
containing a charmonium meson and the other $B$ meson is determined 
to be either a \Bz or \Bzb from its decay products.
The amplitude of the \CP asymmetry, which in the Standard Model 
is proportional to \stwob, is derived from the decay-time distributions in
 such events. We measure $\stwob = 0.741 \pm 0.067\ \stat \pm 0.034\ \syst$ and 
$|\lambda| = 0.948 \pm 0.051\ \stat \pm 0.030\ \syst$.
The magnitude of $\lambda$ is consistent with unity, in agreement with the Standard
Model expectation of no direct \CP violation in these modes.
\end{abstract}


\maketitle
The Standard Model of electroweak interactions describes \CP\ violation
in weak interactions as a consequence of a complex phase in the
three-generation Cabibbo-Kobayashi-Maskawa (CKM) quark-mixing
matrix~\cite{CKM}. In this framework, measurements of \CP  asymmetries in 
the proper-time distribution of neutral $B$ decays to charmonium final states provide
a direct measurement of \stwob~\cite{BCP}, where 
$\beta \equiv \arg \left[\, -V_{\rm cd}^{}V_{\rm cb}^* / V_{\rm td}^{}V_{\rm tb}^*\, \right]$.

Observations of \CP violation in \Bz decays were reported last year by 
the \babar~\cite{babar-stwob-prl} and
Belle~\cite{belle-stwob-prl} collaborations. The \pep2\ collider has since delivered 
an additional 63 \invfb, thereby approximately tripling the data 
sample near the $\Upsilon$(4S) resonance. 
In this Letter we report a more precise measurement of \stwob
using the full sample of about 88 million $B\Bbar$ decays.
The \babar\ detector and the measurement technique 
are described in detail in Refs.~\cite{babar-detector-nim} and~\cite{babar-stwob-prd}, respectively.
Changes in the analysis with respect to the
published result~\cite{babar-stwob-prl} include processing of all data
with a uniform
event reconstruction, a new flavor-tagging algorithm, and the addition of the 
decay mode  $\Bz\to \eta_c \KS$. 

We reconstruct a
sample  of neutral $B$ mesons ($B_{\CP}$) decaying to the final states    
$\jpsi\KS$, $\psitwos\KS$, $\chicone\KS$, $\eta_c \KS$, 
$\jpsi\Kstarz (\Kstarz \to \KS\piz)$, and $\jpsi\KL$. 
The $\jpsi$ and $\psitwos$ mesons are reconstructed through their decays
to $e^+e^-$ and $\mu^+\mu^-$; the $\psitwos$ is also reconstructed
through its decay to $\jpsi\pi^+\pi^-$. We reconstruct $\chicone$ mesons
in the decay mode $\jpsi\gamma$ and $\eta_c$ mesons
in the $\KS K^+\pi^-$ and $K^+K^-\pi^0$ final states~\cite{chargeconj}. 
The $\KS$ is reconstructed in its 
decay to $\pipi$ (and to $\ppz$ for the $\jpsi\KS $ mode).
We examine each event in the $B_{\CP}$ sample for evidence that the recoiling
$B$ meson decayed as a \Bz or \Bzb (flavor tag).

The proper-time distribution of $B$ meson decays to a \CP eigenstate with a \Bz
or \Bzb tag can be expressed in terms of a complex parameter $\lambda$
that depends on both the \Bz-\Bzb oscillation amplitude and the amplitudes
describing \Bzb and \Bz decays to this final
state~\cite{lambda}. The decay rate  ${\rm f}_+({\rm f}_-)$ when the 
tagging meson is a $\Bz (\Bzb)$ is given by 
\begin{eqnarray}
{\rm f}_\pm(\, \deltat) = {\frac{e^{{- \left| \deltat \right|}/\tau_{\Bz} }}{4\tau_{\Bz} }}  
\Bigg[ 1 \Bigg.& \!\!\! \pm& \!\!\!  {\frac{2\mathop{\cal I\mkern -2.0mu\mit m}\lambda}{1 + |\lambda|^2} }
  \sin{( \Delta m_{d}  \deltat )} \nonumber \\
 &\!\!\! \mp& \!\!\! \Bigg. {\frac{1  - |\lambda|^2 } {1 + |\lambda|^2} }
       \cos{( \Delta m_{d}  \deltat) }  \Bigg],
\label{eq:timedist}
\end{eqnarray}
where $\Delta t = t_{\rm rec} - t_{\rm tag}$ is the difference between 
the proper decay times of the reconstructed $B$ meson ($B_{\rm rec}$) and 
the tagging $B$ meson ($B_{\rm tag}$),
$\tau_{\Bz}$ is the \Bz lifetime, and \deltamd is the
\Bz-\Bzb oscillation frequency.
The sine term in Eq.~\ref{eq:timedist} is due to the interference between direct
decay and decay after flavor change, and the cosine term is due to the
interference between two or more decay amplitudes with different weak
and strong phases. \CP violation can be observed as a difference
between the \deltat distributions of \Bz- and \Bzb-tagged events or as
an asymmetry with respect to $\deltat = 0$ for either flavor tag. 
\par
In the Standard Model, $\lambda=\eta_f e^{-2i\beta}$ for
charmonium-containing $b\to\ccbar s$ decays, where $\eta_f$ is the \CP
eigenvalue of the final state $f$. Thus, the time-dependent
\CP asymmetry is
\begin{eqnarray}
A_{\CP}(\deltat) &\equiv&  \frac{ {\rm f}_+(\deltat)  -  {\rm f}_-(\deltat) }
{ {\rm f}_+(\deltat) + {\rm f}_-(\deltat) } \nonumber \\
&=& -\eta_f \stwob \sin{ (\Delta m_{d} \, \deltat )} ,
\label{eq:asymmetry}
\end{eqnarray}
with $\eta_f=-1$ for $\jpsi\KS$, $\psitwos\KS$, $\chicone \KS$, and $\eta_c \KS$, and
$+1$ for $\jpsi\KL$.
Due to the presence of even ($L$=0, 2) and odd ($L$=1) orbital angular momenta in the 
$B\to\jpsi\Kstarz$ final state, there can be \CP -even and \CP -odd contributions to the decay rate. 
When the angular information in the decay is ignored, the measured \CP\ asymmetry 
in $\jpsi\Kstarz$ is reduced by a factor $1-2R_{\perp}$, where
$R_{\perp}$ is the fraction of the $L$=1 component. We have measured  
$R_{\perp} = (16.0 \pm 3.5)\% $~\cite{BABARTRANS}, which 
gives $\eta_f = 0.65 \pm 0.07$ after acceptance corrections in the $\jpsi\Kstarz$ mode.

The event selection, lepton and $K^{\pm}$ identification, and \jpsi
and $\psitwos$ reconstruction used in this analysis are similar to
those described in Ref.~\cite{babar-stwob-prd},
as are the selection criteria for the channels
$\jpsi\KS $, $\psitwos\KS $, $\chicone\KS$, $\jpsi\Kstarz$, and
$\jpsi\KL $.
The $\Bz\to\eta_c\KS$ sample selection is described in Ref.~\cite{etacks}.
In brief, the $K^\pm$ candidates must satisfy kaon identification criteria and the  
$\KS \to \pipi$ and $\pi^0 \to \gamma \gamma$ candidates are required to have reconstructed masses 
within 12.5 and 15~\mevcc, respectively, of their nominal masses~\cite{PDG2000}.
The $\eta_c$ candidates (with $2.90 < M_{KK\pi} < 3.15 \gevcc$)  are 
combined with $\KS \to \pipi$ candidates reconstructed within 10~\mevcc\ of the \KS\ nominal mass
to form a $\Bz$ candidate.
This sample includes a contribution of $(15\pm 2)\%$ from hadronic $\jpsi$ decays to the 
$KK\pi$ final states.
\par
We select candidates in the $\Bz\to\jpsi\KS$, $\psitwos\KS$, $\chicone\KS$, and
$J/\psi K^{*0}$ sample by requiring that the
difference $\Delta E$ between their energy and the beam energy in the
center-of-mass frame be less than three standard deviations from zero.
The $\Delta E$ resolution is about 10\mev, except for
the mode with $\KS\to\piz\piz$ (33\mev) and with $\Kstarz$ (20\mev).
The $\Bz\to\eta_c\KS$ candidates are required to have $|\Delta E|$ less than 40 (70) MeV 
for the  $\KS K^+\pi^-$ ($K^+K^-\pi^0$) modes.
For all modes except $\jpsi\KL$,
the beam-energy substituted mass
$\mes=\sqrt{{(E^{\rm cm}_{\rm beam})^2}-(p_B^{\rm cm})^2}$ must be greater than
$5.2$ $\gevcc$.
To determine numbers of events and purities,
a signal region $5.270\ (5.273) < \mes < 5.290\ (5.288)\gevcc$ is used for 
modes containing $\KS$ ($\Kstarz$). 
In the $\jpsi\KL$ mode, the $\Delta E$ resolution is 3.5 \mev (after \B\ mass constraint)
and  the signal region is defined by $|\Delta E| < 10\mev$.  
\par
A measurement of $A_{\CP}$ requires a determination of the experimental
$\Delta t$ resolution and the fraction $w$ of events in which the tag
assignment is incorrect. This mistag fraction reduces the observed
\CP asymmetry by a factor $1-2w$. 
Mistag fractions and $\Delta t$ resolution functions 
are determined from a sample of neutral $B$ mesons that decay to flavor eigenstates 
($B_{\rm flav}$) consisting of the channels
$D^{(*)-}h^+ (h^+=\pi^+,\rho^+$, and $a_1^+)$ and $\jpsi\Kstarz
(\Kstarz\to\Kp\pim)$.
Validation studies are performed with a control sample of $B^+$
mesons decaying to the final states $\jpsi K^{(*)+}$, $\psitwos
K^+$, $\chicone \Kp$,  $\eta_c \Kp$, and $\overline D^{(*)0}\pip$.
\par
We use multivariate algorithms to identify signatures of $B$ decays that
determine the flavor of $B_{\rm tag}$.
Primary leptons from semileptonic $B$ decays are selected
from identified electrons and muons
as well as isolated energetic tracks.
We use the charges of identified kaon candidates to define a kaon tag.
Soft pions from \Dstarp decays are selected on the basis
of their momentum and direction with respect to
the thrust axis of $B_{\rm tag}$.
A neural network, which combines the outputs of these physics-based algorithms,
takes into account correlations between different sources of
flavor information and provides an estimate of the mistag probability for each event.
\par
By using the outputs of the physics-based algorithms and the estimated mistag probability,
each event is assigned to one of four hierarchical, mutually exclusive tagging categories.
The \leptontag\ category contains events with an identified lepton, and a supporting
kaon tag if present.
Events with a kaon candidate and soft pion with opposite charge and
similar flight direction are assigned to the \kaonitag\ category.
Events with only a kaon tag are assigned to the \kaonitag\ or \kaoniitag\ 
category depending on the estimated mistag probability.
The \kaoniitag\ category also contains the remaining events with a soft pion.
All other events are assigned to the \othertag\
category or excluded from further analysis based
on the estimated mistag probability.
The tagging efficiencies $\eps_i$
for the four tagging categories are measured from data and summarized in·
Table~\ref{tab:mistag}.
The figure of merit for tagging is the effective tagging efficiency
$Q \equiv \sum_i {\eps_i (1-2\mistag_i)^2} $.
This algorithm improves $Q$ by about 7\% (relative) over the algorithm used in
Ref.~\cite{babar-stwob-prd}.
\par
The time interval \deltat between the two $B$ decays is calculated
from the measured separation \deltaz between the decay vertices of
$B_{\rm rec}$ and $B_{\rm tag}$ along the collision ($z$) axis~\cite{babar-stwob-prd}.
We determine the $z$ position of the $B_{\rm rec}$ vertex from
its charged tracks. The $B_{\rm tag}$ decay vertex
is determined by fitting tracks not belonging to the $B_{\rm rec}$ 
candidate to a common  vertex, employing constraints from the beam spot
location and the $B_{\rm rec}$ momentum~\cite{babar-stwob-prd}. 
We accept events with a \deltat\ uncertainty of less than 2.5\ps
and $\vert \deltat \vert <20 \ps$.
The fraction of events satisfying these requirements is 95\%.
The r.m.s. \deltat resolution for 99.7\% of these events is 1.1\ps.
\begin{table}[!t]
\caption
{Efficiencies $\epsilon_i$, average mistag fractions $\mistag_i$, mistag fraction differences
$\Delta\mistag_i=\mistag_i(\Bz)-\mistag_i(\Bzb)$, and $Q$ extracted for each tagging
category $i$ from the $B_{\rm flav}$ and $B_{\CP}$ samples. 
}
\label{tab:mistag} 
\begin{ruledtabular} 
\begin{tabular*}{\hsize}{l
@{\extracolsep{0ptplus1fil}}  D{,}{\ \pm\ }{-1} 
@{\extracolsep{0ptplus1fil}} D{,}{\ \pm\ }{-1} 
@{\extracolsep{0ptplus1fil}}  D{,}{\ \pm\ }{-1} 
@{\extracolsep{0ptplus1fil}}  D{,}{\ \pm\ }{-1}}  
Category     & 
\multicolumn{1}{c}{$\ \ \ \varepsilon$   (\%)} & 
\multicolumn{1}{c}{$\ \ \ \mistag$       (\%)} & 
\multicolumn{1}{c}{$\ \ \ \Delta\mistag$ (\%)} &
\multicolumn{1}{c}{$\ \ \ Q$             (\%)} \\ \colrule  
\leptontag   &  9.1,0.2 &  3.3, 0.6 & -1.5,1.1 &   7.9,0.3  \\  
\kaonitag    & 16.7,0.2 & 10.0, 0.7 & -1.3,1.1 &  10.7,0.4  \\ 
\kaoniitag   & 19.8,0.3 & 20.9, 0.8 & -4.4,1.2 &   6.7,0.4  \\ 
\othertag    & 20.0,0.3 & 31.5, 0.9 & -2.4,1.3 &   2.7,0.3  \\  \colrule 
All          & 65.6,0.5 &           &          &  28.1,0.7  \\ 
\end{tabular*} 
\end{ruledtabular} 
\end{table} 
\begin{figure}[!b]
\begin{center}%
\epsfig{figure=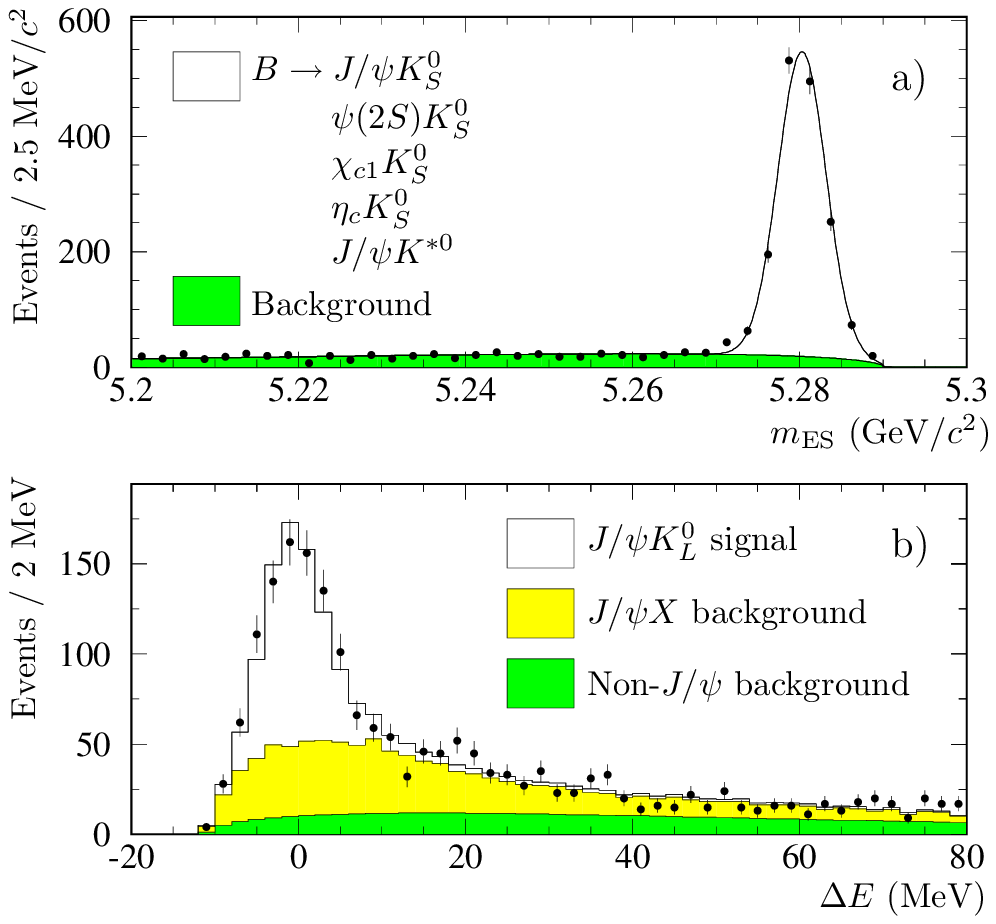, width=0.95\linewidth}
\caption{
Distributions for $B_{\CP}$ candidates satisfying the tagging and vertexing requirements:
a) \mes\ for the final states $J/\psi\KS $, $\psi(2S)\KS$, $\chi_{c1}\KS$,
$\etac\KS$, and $J/\psi K^{*0}(K^{*0}\to \KS\pi^0)$, and
b) $\Delta E$ for the final state $\jpsi\KL$.}
\label{fig:bcpsample}
\end{center}
\end{figure}
\begin{table}[!htb] 
\caption{ 
Number of events $N_{\rm tag}$ in the signal region after tagging and vertexing requirements, 
signal purity $P$,
and results of fitting for \CP\ asymmetries in the $B_{\CP}$ sample and in
various subsamples, as well as in the $B_{\rm flav}$ and charged $B$ control samples.  
Errors are statistical only.}
\label{tab:result} 
\begin{ruledtabular} 
\begin{tabular*}{\hsize}{ l@{\extracolsep{0ptplus1fil}} r c@{\extracolsep{0ptplus1fil}} D{,}{\ \pm\ }{-1} } 
 Sample  & $N_{\rm tag}$ & $P(\%)$ & \multicolumn{1}{c}{$\ \ \ \stwob$}
\\ \colrule
$\jpsi\KS$,$\psitwos\KS$,$\chicone\KS$,$\etac\KS$   & $1506$        & $94$       &  0.76, 0.07   \\
$\jpsi \KL$ $(\eta_f=+1)$                           & $988$        & $55$       &  0.72, 0.16   \\
$\jpsi\Kstarz (\Kstarz \to \KS\piz)$                 & $147$         & $81$       &  0.22, 0.52   \\
\hline
 Full \CP\ sample                                   & $2641$        & $78$       &  0.74,0.07   \\ 
\hline
\hline
\multicolumn{4}{l}{$\jpsi\KS$, $\psitwos\KS$, $\chicone\KS$, $\etac\KS$ only  $(\eta_f=-1)$ }  \\
\hline
$\ \ \jpsi \KS$ ($\KS \to \pi^+ \pi^-$)    & $974$        & $97$       &  0.82, 0.08 \\
$\ \ \jpsi \KS$ ($\KS \to \pi^0 \pi^0$)    & $170$        & $89$       &  0.39, 0.24 \\
$\ \ \psi(2S) \KS$ ($\KS \to \pi^+ \pi^-$) & $150$        & $97$       &  0.69, 0.24 \\
$\ \ \chicone \KS $                        & $80$         & $95$       &  1.01, 0.40 \\
$\ \ \etac\KS $                            & $132$        & $73$       &  0.59, 0.32 \\
\hline
$\ $ \leptontag\ category                & $220$        & $98$       &  0.79, 0.11   \\
$\ $ \kaonitag\ category                 & $400$        & $93$       &  0.78, 0.12   \\
$\ $ \kaoniitag\ category                & $444$        & $93$       &  0.73, 0.17   \\
$\ $ \othertag\ category                 & $442$        & $92$       &  0.45, 0.28   \\
\hline
$\ $ \Bz\ tags                           & $740$        & $94$       &  0.76, 0.10   \\
$\ $ \Bzb\ tags                          & $766$        & $93$       &  0.75, 0.10   \\
\hline\hline
$B_{\rm flav}$ sample                    & $25375$      & $85$       &  0.02, 0.02   \\
\hline 
$B^+$ sample                             & $22160$      & $89$       &  0.02, 0.02   \\
\end{tabular*} 
\end{ruledtabular} 
\end{table}
\par
The signal region contains 2641 events which satisfy the tagging and 
vertexing requirements.
In Table~\ref{tab:result} we list the number of events and the signal
purity for the tagged $B_{\CP}$ candidates. The purities are
determined from fits to the \mes (all \KS\ modes) or
$\Delta E$ (\KL\ mode) distributions in
data, or from Monte Carlo simulation ($\Kstarz$ mode).
Figure~\ref{fig:bcpsample} shows the \mes distribution for modes
containing a $\KS$ or $\Kstarz$ and $\Delta E$ distribution for
the $\jpsi\KL$ candidates. For all modes except $\eta_c \KS$ 
and $\jpsi\KL$, we use simulated events to estimate the fractions of
events in the Gaussian component of the \mes fits
due to cross-feed from other decay modes.
For the $\eta_c\KS$ mode the cross-feed fraction is determined from a fit
to the $M_{KK\pi}$ and \mes\ distributions.
These fractions
range from $(0.3\pm 0.1)$\% for $\jpsi\KS\ (\KS\to\pip\pim)$ to
$(13.1\pm 5.9)$\% for $\eta_c\KS$.
For the $\jpsi\KL$ and $\jpsi\Kstarz$ decay modes, the composition, effective $\eta_f$, and
$\Delta E$ distribution ($\jpsi \KL$ only)
of the individual background sources are
determined either from simulation (for $B\to\jpsi X$) 
or from the $m_{\ell^+ \ell^-}$ sidebands in data (for fake $\jpsi\to \ell^+ \ell^-$).
\par
We determine \stwob with a simultaneous unbinned maximum likelihood fit
to the \deltat distributions of the tagged $B_{\CP}$ and $B_{\rm flav}$
samples. In this fit the \deltat\ distributions of the $B_{\CP}$ sample are described by
Eq.~\ref{eq:timedist} with $|\lambda|=1$.
The \deltat distributions of the $B_{\rm flav}$ sample evolve
according to the known frequency for flavor oscillation in $B^0$
mesons. The observed amplitudes for the \CP asymmetry in the
$B_{\CP}$ sample and for flavor oscillation in the $B_{\rm flav}$ sample
are reduced by the same factor $1-2\mistag$ due to flavor mistags.
Events are assigned signal and background probabilities based on
the \mes\ (all modes except $\jpsi\Kstarz$ and $\jpsi\KL$) or
$\Delta E$ ($\jpsi\KL$) distributions.
The \deltat distributions for the signal are
convolved with a common resolution function, modeled by
the sum of three Gaussians~\cite{babar-stwob-prd}.
Backgrounds are incorporated with an empirical
description of their \deltat spectrum, containing prompt and 
non-prompt components convolved with a resolution
function~\cite{babar-stwob-prd} distinct from that of the signal.
\par
There are 34 free parameters in the fit: \stwob (1),
the average mistag fractions $\mistag$ and the
differences $\Delta\mistag$ between \Bz\ and \Bzb\ mistag fractions for each
tagging category (8), parameters for the signal \deltat resolution (8),
and parameters for background time dependence (6), \deltat resolution
(3), and mistag fractions (8).
We fix $\tau_{\Bz}=1.542\ps$ and $\deltamd
=0.489\ps^{-1}$~\cite{PDG2000}.
The determination of the mistag fractions and \deltat resolution
function parameters for the signal is dominated by the high-statistics $B_{\rm flav}$ sample. 
The measured mistag fractions are listed in Table~\ref{tab:mistag}.
Background parameters are determined from events with
$\mes < 5.27\gevcc$ (except $\jpsi\KL$ and $\jpsi\Kstarz$).
The largest correlation between \stwob\ and any linear combination of the
other free parameters is 0.13. 
We observe a bias of $0.014 \pm 0.005$ in the fitted value of \stwob in
simulated events. Part of this bias ($0.004$) is due to a correlation
between the mistag fractions and the $\Delta t$ resolution not explicitly
incorporated in the fit. Therefore we subtract $0.014$ from the fitted value of
\stwob in data and include $0.010$ in the systematic error.
\par
\begin{figure}[ht]
\begin{center}
 \epsfig{figure=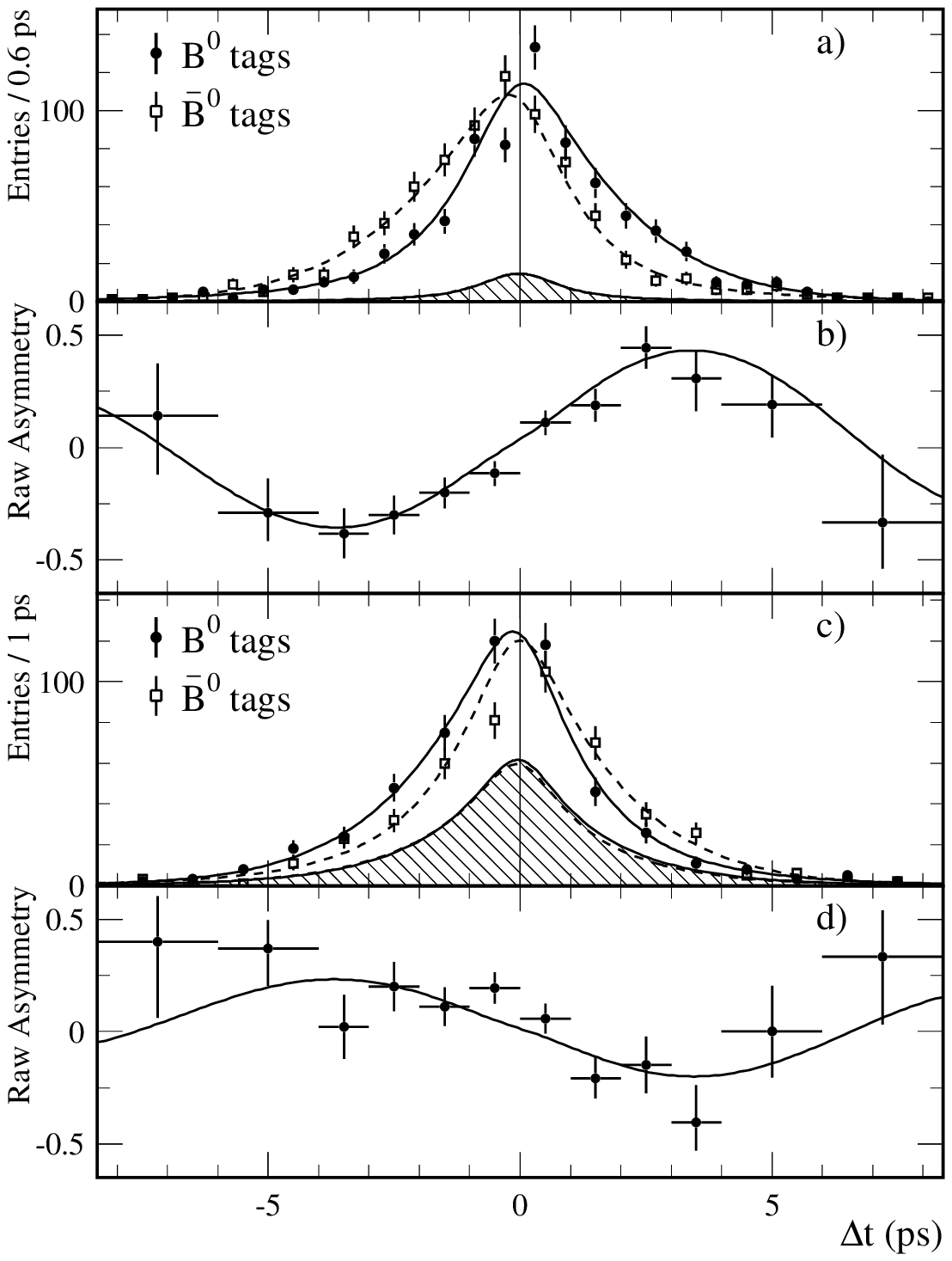,width=1.00\linewidth} 
\caption{
a) Number of $\eta_f=-1$ candidates 
($J/\psi \KS$,
$\psi(2S) \KS$,
$\chicone \KS$, and
$\eta_c \KS$)
in the signal region with a \Bz tag $N_{\Bz }$ and
with a \Bzb tag $N_{\Bzb}$, and b) the raw asymmetry
$(N_{\Bz}-N_{\Bzb})/(N_{\Bz}+N_{\Bzb})$ as functions of \deltat.
The solid (dashed) curves represent the fit projection
in \deltat for \Bz (\Bzb) tags.
The shaded regions represent the background contributions.
Figures c) and d) contain the corresponding information for the $\eta_f=+1$
mode $J/\psi \KL$.
}
\label{fig:cpdeltat}
\end{center}
\end{figure}

The fit to the $B_{\CP}$ and $B_{\rm flav}$ samples yields
\begin{eqnarray}
\stwob=0.741 \pm 0.067\ \stat \pm 0.034\ \syst.\nonumber
\end{eqnarray}
\noindent
Figure~\ref{fig:cpdeltat} shows the \deltat distributions and 
asymmetries in yields between \Bz tags and \Bzb tags for the
$\eta_f=-1$ and $\eta_f = +1$ samples as a function of \deltat,
overlaid with the projection of the likelihood fit result.
\par
The dominant sources of systematic error are 
the uncertainties in the level, composition, and \CP\ asymmetry of
the background in the selected \CP events (0.023),
the assumed parameterization of the \deltat\ resolution function (0.017),
due in part to residual uncertainties in the internal alignment of
the vertex detector, and
possible differences between the $B_{\rm flav}$ and $B_{CP}$ mistag 
fractions (0.012).
The total systematic error is $0.034$.
Most systematic  errors are determined with data and will continue to decrease 
with additional statistics.
\par
The large $B_{\CP}$ sample allows a number of consistency
checks, including separation of the data by decay mode, tagging category,
and $B_{\rm tag}$ flavor. The results of fits 
to these $\eta_f=-1$ subsamples are shown in
Table~\ref{tab:result} and found to be statistically consistent.
The results of fits to the control samples of non-\CP decay modes
indicate no statistically significant asymmetry.
\par
We also measure the parameter $\vert\lambda\vert$ in Eq.~\ref{eq:timedist}
from a fit to the $\eta_f=-1$ sample, which has high
purity and requires minimal assumptions on the effect of backgrounds.
This parameter is sensitive to the difference in the number
of \Bz- and \Bzb-tagged events. In order to account for differences
in reconstruction and tagging efficiencies for \Bz\ and \Bzb mesons,
we incorporate five additional free parameters
in this fit. We obtain $\vert\lambda\vert = 0.948 \pm 0.051\ \stat \pm
0.030\ \syst$. The coefficient of the $\sin(\deltamd \deltat)$ term in Eq.~\ref{eq:timedist} is
measured to be $0.759\pm 0.074\ \stat$.
  The dominant contribution to the systematic error for $\vert\lambda\vert$,
  conservatively estimated to be 0.025, is due
  to interference between the suppressed $\bar b\to \bar u c \bar d$ amplitude with
  the favored $b\to c \bar u d$ amplitude for some tag-side $B$ decays.
The other sources of systematic error for $\vert\lambda\vert$ are the same as in the
\stwob measurement.
\par
This measurement of $\stwob$ supersedes our previous result~\cite{babar-stwob-prl} 
and improves upon the precision of each of the previous measurements~\cite{babar-stwob-prl,belle-stwob-prl} 
by a factor of two.
While the  measured value is consistent with the range implied by the 
measurements and theoretical estimates of the 
magnitudes of CKM matrix elements in the context of the Standard Model, it provides a precise and 
model-independent constraint on the position of the apex of the Unitarity Triangle~\cite{CKMconstraints}.  

We are grateful for the excellent luminosity and machine conditions
provided by our \pep2\ colleagues, 
and for the substantial dedicated effort from
the computing organizations that support \babar.
The collaborating institutions wish to thank 
SLAC for its support and kind hospitality. 
This work is supported by
DOE
and NSF (USA),
NSERC (Canada),
IHEP (China),
CEA and
CNRS-IN2P3
(France),
BMBF and DFG
(Germany),
INFN (Italy),
NFR (Norway),
MIST (Russia), and
PPARC (United Kingdom). 
Individuals have received support from the 
A.~P.~Sloan Foundation, 
Research Corporation,
and Alexander von Humboldt Foundation.


\begin{thebibliography}{99}
\bibitem{chargeconj}
Charge conjugation is implied throughout this letter, unless explicitly stated.
\end{thebibliography}
\end{document}